\documentclass[a4paper]{article}
\usepackage{setspace}
\usepackage[pages=all, color=black, position={current page.south}, placement=bottom, scale=1, opacity=1, vshift=5mm]{background}
\SetBgContents{
 
}      

\usepackage[margin=1in]{geometry} 

\usepackage{amsmath}
\usepackage{amsthm}
\usepackage{amssymb}
\usepackage{multirow}

\usepackage[utf8]{inputenc}
\usepackage{hyperref}
\hypersetup{
	unicode,
}

\usepackage[sort&compress,numbers,square]{natbib}
\theoremstyle{plain}

\theoremstyle{definition}

\usepackage{graphicx, color}
\graphicspath{{fig/}}

\usepackage{algorithm, algpseudocode} 
\usepackage{mathrsfs} 

\usepackage{lipsum}

\usepackage{color}

\title{Bridging the Digital Divide: Mapping Internet Connectivity Evolution, Inequalities, and Resilience in six Brazilian Cities}

\author{Nicol\`o Gozzi$^{1,2}$, Niccol\`o Comini$^{2}$, Nicola Perra$^{3,2,4}$}

\date{
    $^1$ ISI Foundation, Turin, Italy\\
    $^2$ The World Bank Group\\
    $^3$ School of Mathematical Sciences, Queen Mary University of London, UK\\
    $^4$ The Alan Turing Institute, London, UK
	\vspace{0.5cm}
}

\usepackage{url}

\begin{document}
	\maketitle

\begin{abstract}
We investigate the evolution of Internet speed and its implications for access to key digital services, as well as the resilience of the network during crises, focusing on six major Brazilian cities: Belo Horizonte, Brasília, Fortaleza, Manaus, Rio de Janeiro, and São Paulo. Leveraging a unique dataset of Internet Speedtest\textsuperscript{\tiny\textregistered} results provided by Ookla\textsuperscript{\tiny\textregistered}, we analyze Internet speed trends from $2017$ to $2023$. Our findings reveal significant improvements in Internet speed across all cities. However, we find that prosperous areas generally exhibit better Internet access, and that the dependence of Internet quality on wealth have increased over time. Additionally, we investigate the impact of Internet quality on access to critical online services, focusing on e-learning. Our analysis shows that nearly $13\%$ of catchment areas around educational facilities have Internet speeds below the threshold required for e-learning, with less rich areas experiencing more significant challenges. Moreover, we investigate the network's resilience during the COVID-19 pandemic, finding a sharp decline in network quality following the declaration of national emergency. We also find that less wealthy areas experience larger drops in network quality during crises. Overall, this study underscores the importance of addressing disparities in Internet access to ensure equitable access to digital services and enhance network resilience during crises.
\end{abstract}

\maketitle

\section{Introduction}

The widespread availability of Internet connectivity has transformed several aspects of our lives. From communication and commerce to education and entertainment, access to reliable, fast, and affordable Internet connectivity has become a critical factor for promoting economic and social development~\cite{mora2021internet, medeiros2021infrastructure, galperin2017connected}. Despite the large overall improvements in technology and adoption witnessed over the last decades, we still observe huge gaps in access to digital services and varying levels of digital literacy. The COVID-19 pandemic has shown the impact of such digital divide and highlighted the importance of addressing it. Indeed, during the acute phases of the crisis, as numerous activities rapidly migrated online, unequal access to a reliable Internet connection affected the possibility to carry out activities from home increasing the possible exposures to the virus for the unconnected ~\cite{covid19_digitalinfra, NBERw26982,mariscal2020impact, gozzi2023adoption}. Particularly clear are the negative impact of Internet connectivity disparities on educational achievement, access to tele-medicine, and adoption of remote working~\cite{torres2020transition,Bauer2020OvercomingMH,taddei2020facing,work_digitaldivide,Soomro2020,AZUBUIKE2021100022,Eruchalu2021,Watts2020,taylor2021decreasing}

In this context, we aim to investigate how Internet connectivity has evolved over the past years across regions and socio-economic strata, its impact on the access to key services, and its resilience to extraordinary events such as the COVID-19 Pandemic. As a case study, we consider six major Brazilian cities: Belo Horizonte, Brasília, Fortaleza, Manaus, Rio de Janeiro, and São Paulo. Brazil reports one of the highest GINI index in the world~\cite{giniWB} and inequality has been one of the main issues affecting its socio-economic development for decades. On the other hand, Brazil can compete with the most advanced areas in the world when it comes to digital capabilities. Indeed, it hosts cloud services of some of the most important providers, and it is home to several high-tech startups. However, the inequality observed in the socioeconomic dimension, is also reflected in the digital sector. For instance, while the overall average broadband fixed access for every $100$ inhabitants is $24$~\cite{anatel} there is a significant heterogeneity among states. Some Brazilian States such as Santa Catarina ($36.15$) outperform OECD countries (e.g., Italy, $32.1$) while others such as Acre ($13.8$), Amazonas ($13.8$), and Maranhao ($9.9$) report remarkably lower figures. Additionally, access and usage of digital tools is far from being inclusive and several areas, even within wealthier states, face a dramatic digital inequality. For these reasons, Brazil constitutes a perfect representation of the complex socio-economic dynamics and challenges that public and private sector face in addressing the digital gap.

To quantify Internet quality and its evolution in these cities, we leverage a unique dataset provided by Ookla consisting of nearly $100M$ geolocalized Speedtest results, collected in the time window spanning from $2017$ to $2023$. We split the analysis in two parts. In the first, we focus on characterising the spatio-temporal evolution of Internet connectivity by exploring differences across socio-economic indicators. In the second part instead, we study Internet connectivity indicators in the catchment areas of educational activities and quantify the resilience of the digital infrastructure during the COVID-19 Pandemic.

We find significant improvements in Internet quality across all cities considered between $2017$ and $2023$. Interestingly, we observe a trend towards a more homogeneous distribution of Internet speed, indicating reduced dispersion over the years. However, despite this increased homogeneity, we find an increasing correlation between Internet speed and wealth, with wealthier areas experiencing better Internet access and with this gap widening over time. Furthermore, we also find a noticeable increase in spatial autocorrelation of Internet quality over the years, with the emergence of clusters characterized by high and low speeds.

Furthermore, our analysis reveals that approximately $13\%$ of catchment areas around education facilities, where $8\%$ of the school-age population resides, experience Internet speeds insufficient for accessing key digital services such as e-learning. Additionally, these areas tend to exhibit lower wealth, suggesting a compounding effect of inequality.

Finally, we assess the impact of the stress placed on the network following the declaration of the COVID-19 national emergency in Brazil. We find that, on average, this caused a $-20\%$ in download speed across all cities, with values ranging from $-7\%$ in Brasília to almost $-30\%$ in Manaus. Our findings indicate that this impact was more pronounced in less wealthy areas compared to more wealthy ones.

Overall, this study shows that while the evolution of Internet quality showed an overall progress, disparities persist, with socio-economic factors playing significant roles. Addressing these disparities is crucial to ensure equitable access to digital services and to enhance network resilience in times of crisis. This study demonstrates that despite the resources allocated by the public and private sector to the strengthening of the Brazilian digital infrastructure, investments are still needed, particularly in the less affluent areas.

\section{Results}

\subsection{Internet speed evolution analysis}

As a first step, our research aims to analyze the evolution of Internet quality, specifically measured by fixed download speed, across six major Brazilian cities: Belo Horizonte, Brasília, Fortaleza, Manaus, Rio de Janeiro, and São Paulo. To accomplish this, we leverage a unique dataset consisting of $\sim100M$ Internet Speedtest results provided by Ookla. The data covers the period between $2017$ and $2023$. Furthermore, it is geolocalized and provides the download/upload speed (i.e., Megabits per second) and latency in milliseconds for fixed networks. In the Supplementary Information we show results considering mobile network, which we also discuss below.

It is important to highlight from the start how the data serves only as a proxy of Internet quality. Indeed, due to the details of the software/tool used to make a measurement, possible bottlenecks in home networks (e.g., routers), the number of devices connected to a specific network, and selection biases (e.g., tests might be done when users are experiencing connectivity issues or when users need to connect in a new location and/or by more digitally aware users) the outcome of tests might differ from the real Internet speed~\cite{feamster2020measuring,gozzi2023adoption}. Nevertheless, Ookla is the \emph{canonical} network performance testing service. It is widely used to infer the features of Internet connectivity across and within regions by academic and official institutions ~\cite{vakataki2021visualizing,ford2021form,feamster2020measuring}. Furthermore, as described below, our analysis aggregates Speedtest results within specific geographical cells thus averaging among many measurements. This allows to reduce the possible impact of the more technical issues mentioned.

To ensure uniform spatial coverage we partition the geographical area of each city into hexagonal cells, creating a regular grid (see Fig.~\ref{fig:fig5}A). Then, we calculate the Internet speed within each of these units as function of time. This approach allows to explore different resolutions and finer scales with respect to administrative partitions. We also compute a proxy measure for wealth in each of these unit using the Relative Wealth Index (RWI) provided by Meta~\cite{chi2022microestimates}. For a more detailed description of our methodology, please refer to Section~\ref{sec:mem}.

\begin{figure}[ht!]
\centering
\includegraphics[width=\textwidth]{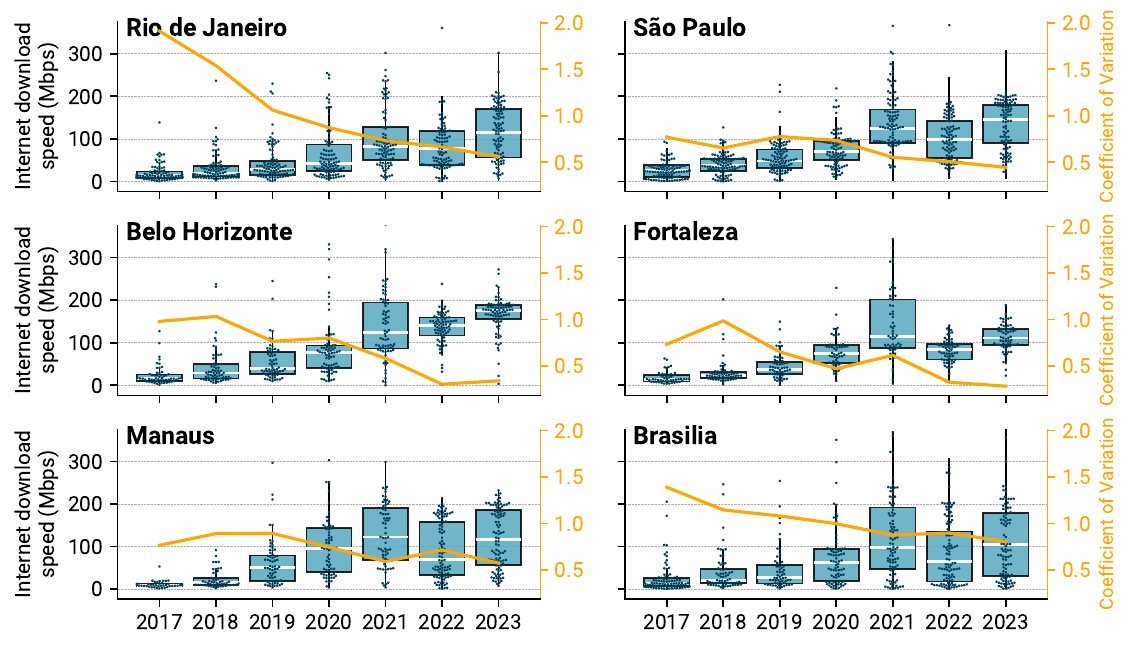}
\caption{Evolution of Internet speed, expressed in download speed (Mbps), across the six cities considered. Boxplots show the distribution of Internet speed within each hexagonal unit in each city. The orange line indicates the evolution of the coefficient of variation of the Internet speed distribution over the years.}
\label{fig:fig1A}
\end{figure}

Figure~\ref{fig:fig1A} shows the evolution of Internet speed, from $2017$ to $2023$ in the six cities. Across the board, our analysis reveals a significant improvement in Internet speed throughout all cities over the past six years. Specifically, Belo Horizonte exhibits the highest median download speed ($176mbps$) in $2023$, followed by São Paulo ($146mpbs$), Manaus ($116mbps$), Rio de Janeiro ($114mbps$) Fortaleza ($111mbps$), and Brasília ($105mbps$). On the other hand, Manaus experienced the highest growth during the period, marking a $+1200\%$ increase, followed by Belo Horizonte ($+1012\%$), Rio de Janeiro ($+719\%$), Fortaleza ($+685\%$), Brasília ($+677\%$), and São Paulo ($+463\%$). Furthermore, in the same plot we show the coefficient of variation of the distribution of Internet speed within each city across the years. The coefficient of variation is a measure of dispersion defined as the ratio between standard deviation and average of a statistical distribution. Our findings indicate a decreasing trend in the coefficient of variation across the six cities, suggesting a persistent trend towards a more homogeneous distribution of Internet speed as it improved. However, we acknowledge differences among the cities examined. Brasília exhibits the highest dispersion in Internet speed distribution in $2023$ ($CV=0.80$), while Fortaleza demonstrates the less disperse distribution ($CV=0.28$). More quantitatively, in $2023$, the ratio between the $3^{rd}$ and $1^{st}$ quartiles of Internet speed is $5.8$ in Brasília, whereas it is only $1.4$ in Fortaleza.

\begin{figure}[ht!]
\centering
\includegraphics[width=\textwidth]{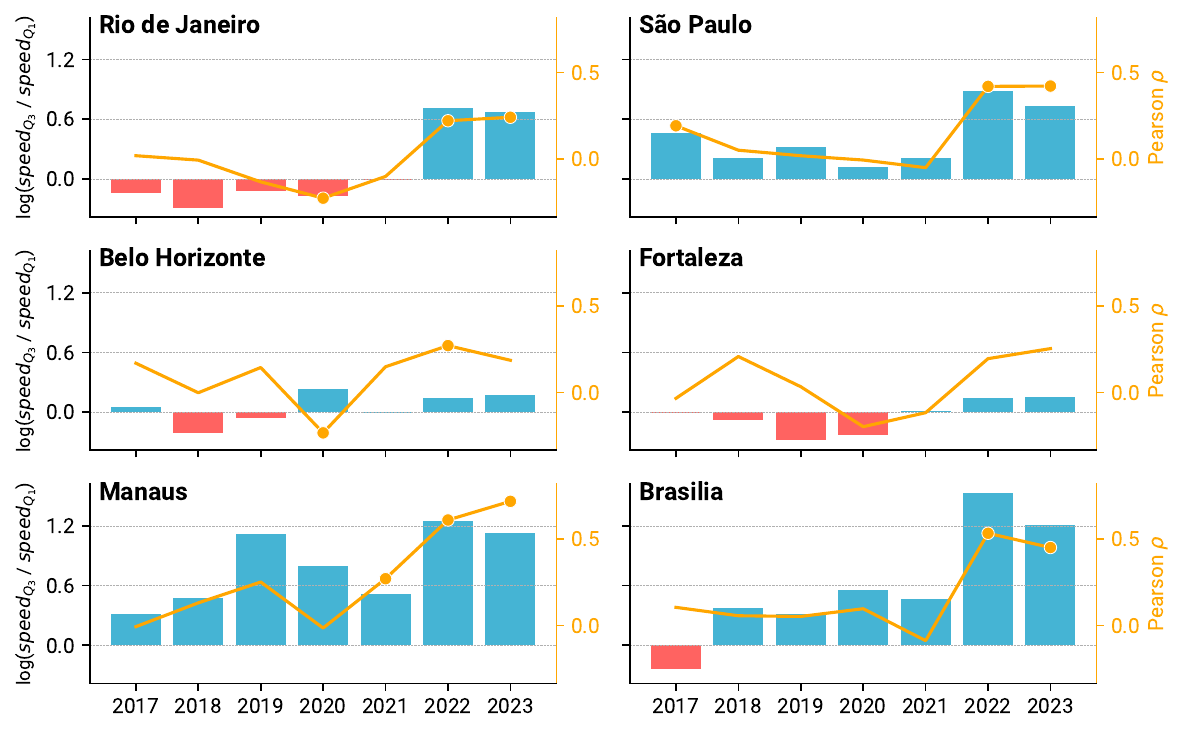}
\caption{Logarithm of the ratio between Internet speeds measured in spatial units with wealth higher than the $75^{th}$ quantile and those with wealth lower than the $25^{th}$ quantile. The orange line represents the Pearson correlation coefficient between Internet speed and RWI of different spatial units. Circles indicate where the coefficient is significant at the $5\%$ level.}
\label{fig:fig1B}
\end{figure}

It is important to highlight how, despite a general trend towards an homogenisation of the dispersion of Internet speeds, the data still reveals persistent and even increasing disparities across socio-economic strata. Figure~\ref{fig:fig1B} shows the logarithm of the ratio between the average Internet speed measured in cells with wealth higher than the $75^{th}$ quantile and those with wealth lower than the $25^{th}$ quantile. This metric is meant to compare and highlight the differences between the wealthiest and the poorest units. A value close to zero indicates similar Internet quality for both wealthy and less wealthy areas, while positive (negative) values denote better Internet quality for the more wealthy (less wealthy). As detailed in Section~\ref{sec:mem}, the wealth of each unit is calculated using the Relative Wealth Index (RWI) provided by Meta~\cite{chi2022microestimates}. 

Across various years and cities, our analysis reveals a consistent trend: wealthy areas generally experience better Internet quality compared to less wealthy areas. Moreover, we note an increasing disparity in Internet quality between more and less wealthy areas over the years. In the case of Manaus and Sao Paolo cells characterized by higher RWI features better Internet quality across the whole time horizon under study. This trend is observed also in Brasilia with the exception of $2017$. In Rio de Janeiro instead, only in the last two years Internet quality in wealthy cells was better with respect to less wealthy areas, though the negative values are closer to zero. Finally in Belo Horizonte and Fortaleza, the values are overall smaller with respect to the other cities though positive in the last years. The association between Internet quality and RWI is supported by the Pearson correlation coefficient between Internet speed and RWI, shown in Figure~\ref{fig:fig1B}. The coefficient has increased across all cities in recent years, with all cities showing a positive correlation as of $2023$, which is significant at $5\%$ level with the exception of Belo Horizonte and Fortaleza.

In the Supplementary Information we repeat the analyses presented in Figure~\ref{fig:fig1A} and Figure~\ref{fig:fig1B} for mobile network. Also in that case, we find a significant overall improvement of mobile network speed over the period considered. Interestingly, we find that, while mobile speed and wealth are also positively correlated, the observed trend in time is decreasing, contrasting the finding in the case of fixed network. 

In the case of Rio de Janeiro, we extend our analysis to include tests conducted both inside and outside \textit{favelas}. The results of this analysis are presented in the Supplementary Information. \textit{Favelas} are informal, densely populated urban settlements in Brazil, typically characterized by substandard housing and a lack of basic services, arising from socio-economic disparities and rapid urbanization. Not surprisingly, we find that tests performed within a \textit{favela} generally exhibit lower internet speeds. Additionally, this disparity has increased over the years. In $2017$, the median speed of tests conducted inside and outside \textit{favelas} was $13.7$ Mbps and $14.3$ Mbps, respectively, reflecting a $4\%$ difference. By $2023$, these speeds had changed to $40.1$ Mbps and $94.2$ Mbps, respectively, resulting in a $57.4\%$ difference.

To investigate whether Internet speed has become more spatially autocorrelated over time, we calculate the Moran's $I$ statistic for download speed in each hexagonal unit across various cities for each year within the study period~\cite{Zhou2008}. The Moran's $I$ quantifies the degree of spatial autocorrelation of a quantity, indicating the extent to which similar values cluster or disperse across geographical units. More in detail, a positive (negative) Moran's $I$ indicates spatial autocorrelation (dispersion) in the dataset, meaning that similar (dissimilar) values tend to cluster together in space. Our analysis reveals the emergence of spatial clusters characterized by high or low Internet speed. This finding is exemplified in Figure~\ref{fig:fig2}A where we present the results for Rio de Janeiro in $2017$, $2020$, and $2023$ (in the Supplementary Information we show results also for other cities). The global Moran's $I$ values exhibit a notable increase from approximately $0$ in $2017$ to $0.17$ in $2020$ and further to $0.40$ in $2023$. Visual inspection of the maps also reveals the emergence of spatial clusters of high and low Internet speed over the years. Specifically, the maps indicate units where the local Moran's $I$ statistic — measuring the spatial clustering pattern of individual observations — is significant at the $5\%$ level, with units colored to denote low-speed (red) or high-speed (blue) clusters. Furthermore, we analyze the evolution of the global Moran's $I$ across different cities over the six-year period. The findings observed in the case of Rio de Janeiro are consistent across cities, with the statistic generally demonstrating an increase over the years. In more details, we observe how in all cities, with the exception of Belo Horizonte, the last two years show the highest values of Moran's $I$. Also, we note how in Brasilia, Fortaleza, and Sao Paulo, the global Moran's $I$, measured considering data collected in $2023$, is smaller with respect to the previous year. The decreasing trend in the last year is also observed, though to a lesser extent, also in the case of Manus and in Belo Horizonte, though in the latter the value obtained it is not significant at the $5\%$ level. In the Supplementary Information we repeat this analysis for mobile network. We find also in that case positive and significant spatial autocorrelation of mobile Internet speed, event tough in this case the temporal trend is less clear.   

\begin{figure}[ht!]
\centering
\includegraphics[width=\textwidth]{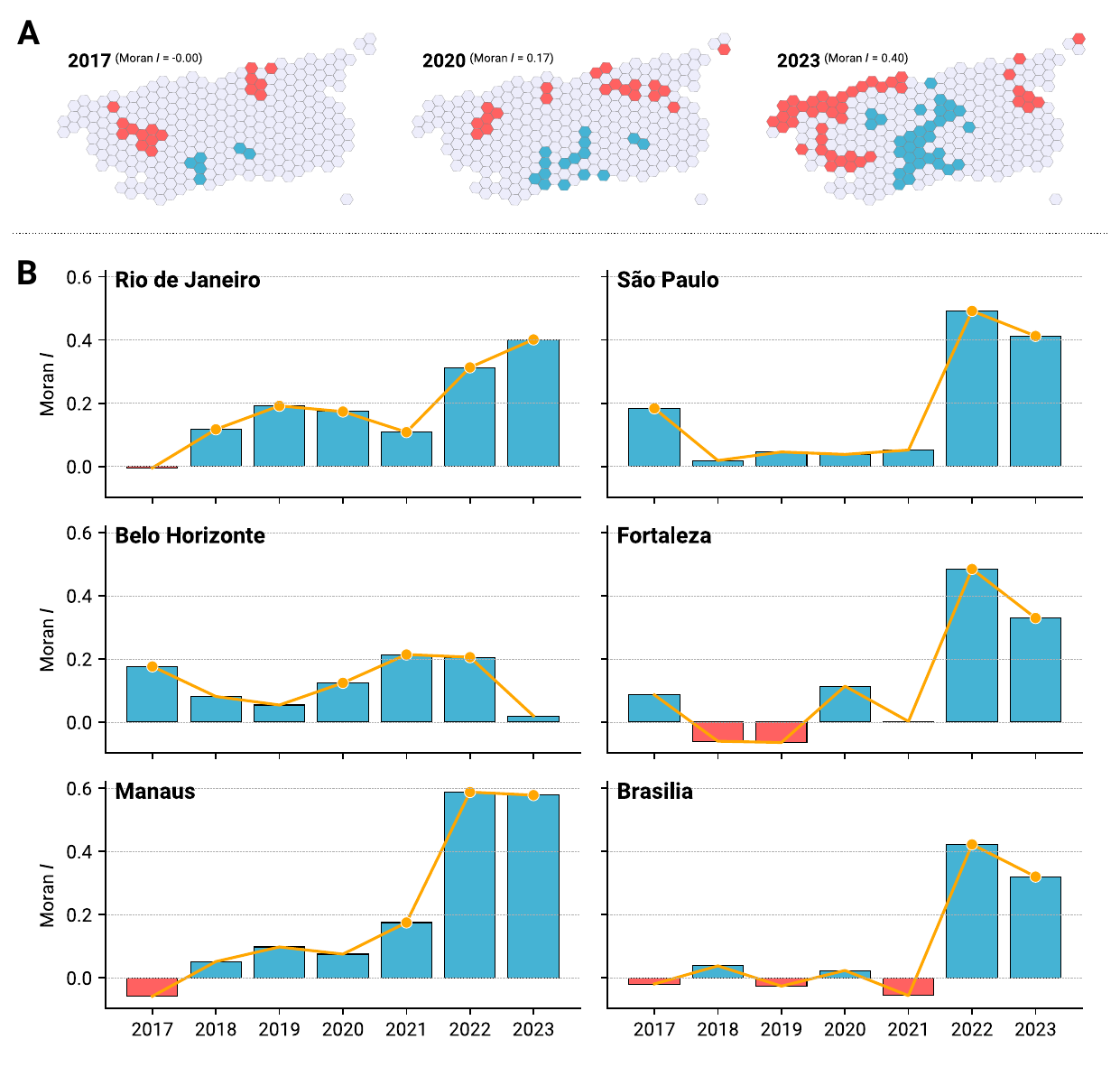}
\caption{\textbf{Spatial Clustering of Internet Speed.} \textbf{A)} Distribution of spatial units with significant local Moran's $I$ in Rio de Janeiro in $2017$, $2020$, and $2023$. Clusters of low (high) Internet speed are shown in red (blue). \textbf{B)} Evolution of global Moran's $I$ in each city between $2017$ and $2023$. Circles indicate where the statistic is significant at the $5\%$ level.}
\label{fig:fig2}
\end{figure}

\subsection{Access to e-learning}

\begin{figure}[ht!]
\centering
\includegraphics[width=\textwidth]{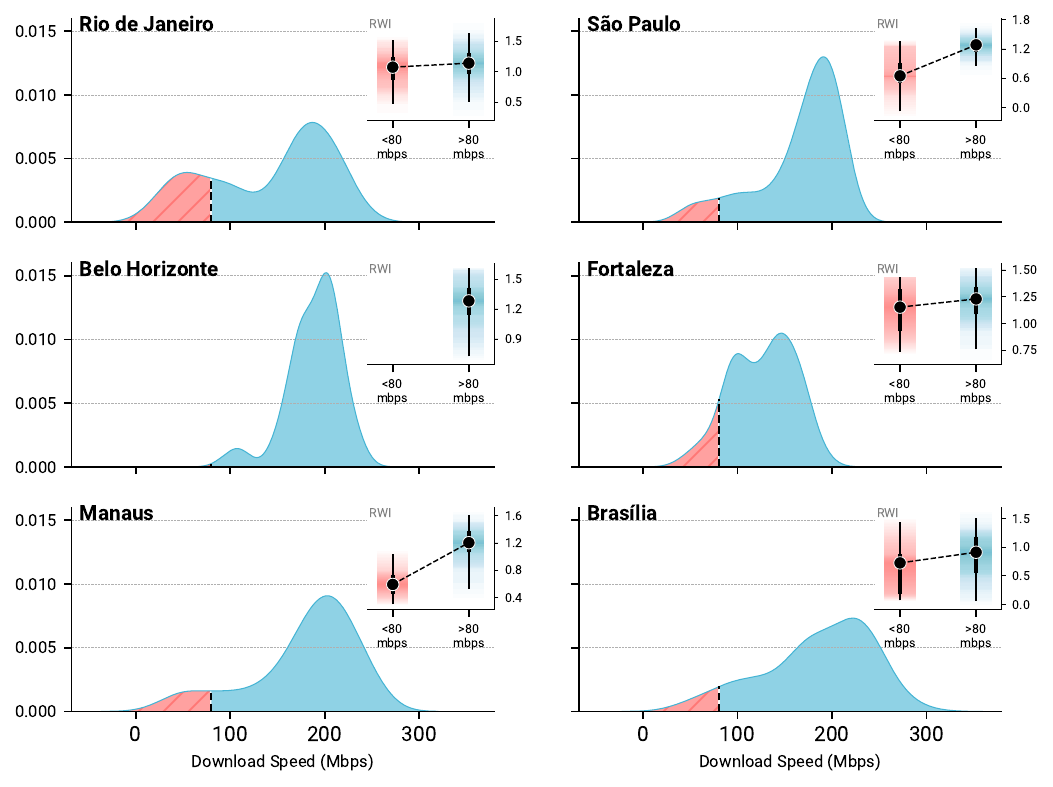}
\caption{\textbf{Internet Speed in Catchment Areas of Education Facilities}. Distribution of Internet speed, measured as download speed (Mbps), in catchment areas of education facilities across all cities (2023). The portion of the distribution where speed is lower than $80$ Mbps is colored in red. In the inset of each plot, RWI distribution of catchment areas of education facilities featuring a download speed above and below $80$ Mbps is shown.}
\label{fig:fig3}
\end{figure}

In the second part of our analysis we investigate whether the disparities in Internet quality highlighted in the previous section may impact access to key services. While acknowledging the diversity and variety of these, in the following we use e-learning as a concrete and arguably important example. Indeed, as mentioned in the Introduction, extant research has highlighted the positive relationship between Internet quality and educational attainment~\cite{Bauer2020OvercomingMH}. Our approach is as follows. First, we gather the locations of educational facilities across the six cities under consideration using data from OpenStreetMap~\cite{hostm_edu}. Next, we conduct a Voronoi tessellation for each city, with the positions of educational facilities as centroids. This process allows us to obtain the catchment area of each educational facility. By construction a catchment area describes the closest educational entity for people living in that region. Subsequently, we compute the Internet speed within each catchment area by aggregating the download speeds of all tests performed within. For this analysis, we consider the most recent data from $2023$. Our aim is to focus solely on recent data to accurately characterize the current disparities in access to essential services. Additionally, we calculate the RWI for each area. Further details on our methodology are available in Section~\ref{sec:mem}. In Figure~\ref{fig:fig3}, we present the distribution of fixed download speeds across all catchment areas in the six cities. We also highlight a threshold of $80$ Mbps (approximately $10$ megabytes per second) as the minimum speed required to access e-learning services~\cite{fcc_bb}. Remarkably, across all cities we find that nearly $13\%$ of catchment areas have speeds below this threshold, affecting approximately $8\%$ of the population in school age. This implies that less than one in every ten children may encounter challenges in accessing e-learning services. We note how e-learning is a general term referring to both synchronous and asynchronous learning  
activities. These span from access to dedicated platforms to ability of exploring broader online resources for homeworks. Nonetheless, we observe a significant variability across cities. In Belo Horizonte, none of the catchment areas exhibit an Internet speed below the $80mbps$ threshold. Following closely is Brasília, with only $4.8\%$ falling below, then São Paulo ($6.8\%$), Fortaleza ($7.4\%$), and Manaus ($8.3\%$). In stark contrast, nearly $24\%$ of catchment areas in Rio de Janeiro fall below this threshold, with approximately $26\%$ of the school-age population residing in these areas.

Additionally, in the inset of each plot in Figure~\ref{fig:fig3}, we display the RWI distribution of catchment areas below and above the $80$ Mbps threshold. Across all cities, our analysis indicates that, on average, catchment areas below the threshold are $15\%$ less wealthy than areas above the threshold. We assess the differences in RWI distribution between the two cases using a t-test, finding a significant difference in the case of Rio de Janeiro, São Paulo, and Manaus (significance level $5\%$). This observation points to a compounding effect of inequality. Indeed, students facing higher challenges in accessing key digital services such as e-learning may already be foreclosed from other  opportunities due to their socio-economic disadvantage.

\subsection{Network resilience during crises}

\begin{figure}[ht!]
\centering
\includegraphics[width=\textwidth]{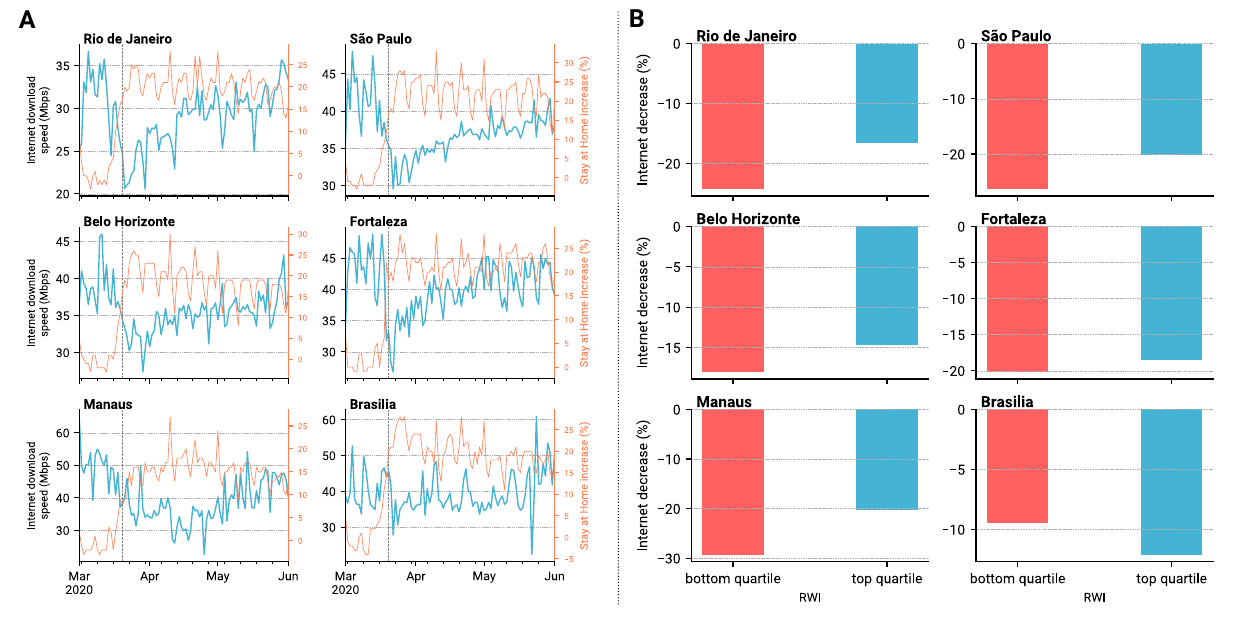}
\caption{\textbf{Network Resilience During the COVID-19 Pandemic}. \textbf{A)} Daily median download speed in the six cities between March and June $2020$. The vertical dashed line indicates when Brazil declared the national emergency. The percentage change in individuals staying at home as measured via Google Community Mobility Reports is also shown. \textbf{B)} Drop in Internet speed following the national emergency declaration in the top and bottom quartiles of the RWI in each city.}
\label{fig:fig4}
\end{figure}

Finally, we aim to investigate the resilience of the network to external shocks and the potential heterogeneous impacts of such events. As a case study, we consider the COVID-19 pandemic. With infections and deaths surging worldwide and restrictions being imposed, the world moved online to maintain essential activities. Arguably, such an unprecedented surge in demand may have affected network quality. In Figure~\ref{fig:fig4}A, we present the median daily download speeds in the six cities between March and June $2020$. Additionally, we mark the date when Brazil declared a national emergency with a vertical dashed line and we show the increase in the percentage of individuals staying at home measured using data from the COVID-19 Community Mobility Reports published by Google~\cite{googleMob}. Across all cities, we observe a sharp decline in network quality, as measured by download speed, following the declaration of the national emergency. Concurrently, the fraction of population staying at home increased. After the initial drop, we observe a gradual recovery, with download speeds approaching pre-emergency levels by June $2020$. Among the cities considered, Manaus experienced the most significant drop in median download speed computed in periods March $1^{st}$-March $20^{th}$ and  March $20^{th}$-April $1^{st}$, with a decline of $-29\%$, while Brasília showed the lowest drop at $-7\%$. The other cities experienced declines ranging from Rio de Janeiro ($-25\%$), Fortaleza ($-21\%$), São Paulo ($-19\%$), to Belo Horizonte ($-16\%$).

Furthermore, in Figure~\ref{fig:fig4}B, we illustrate these drops for the top and bottom quartiles of the RWI. We observe that, with the exception of Brasília, more wealthy areas experienced smaller drops compared to less wealthy areas. When combined with the previous findings, this suggests that besides experiencing slower Internet speeds, less wealthy areas may also face more significant drawbacks during extraordinary stress on the network.

In the Supplementary Information we repeat this analysis for mobile network. Also in that case we find that mobile Internet speed was significantly affected by the stress put on the network following national emergency declaration, even tough we do not observe a clear divide in the drops experienced by more and less wealthy areas.

\section{Discussion}

In this study, we analysed the spatio-temporal evolution of Internet speed in six Brazilian cities spanning the years $2017$ and $2023$. Our analysis revealed a significant increase in Internet speed across all cities, along with a trend towards more uniform distribution. However, we also identified the emergence of spatial clusters characterized by high/low Internet speed. Furthermore, we found an increasing correlation between Internet speed and measures of wealth, indicating that more wealthy areas tended to experience higher Internet speeds over time. Such inequality pattern was also reported by the analysis done in the case of the \textit{favelas} in Rio de Janeiro, which revealed an increasing internet speed gap with the rest of the city.

To further characterize the impact of such disparities, we considered two case studies. In the first one, we focused on the Internet speed in catchment areas around educational facilities in each city. Notably, we find that, as of $2023$, approximately $13\%$ of these areas may have encountered challenges in accessing key digital services such as e-learning. We observed significant variations among cities, with Rio de Janeiro reaching a peak of $24\%$ of these areas falling below the threshold for e-learning. Additionally, we showed that these areas tend to be less wealthy, suggesting a potential compounding effect of inequality, where regions already facing limited access to opportunities may also encounter challenges in digital access.

In our second case study, we examined the unprecedented stress placed on the network due to the shift online driven by the COVID-19 pandemic. Our analysis showed a significant decrease in Internet speed across all cities following the declaration of national emergency. Moreover, we found that less wealthy areas generally experienced more pronounced declines in Internet connectivity during the early weeks of the COVID-19 crisis. This result is even more concerning when combined with findings from a recent study that has shown how access to a fast Internet is an effective measure in case of exogenous shocks such as the pandemic to limit the exposure to infections~\cite{gozzi2023adoption}. 

Overall, these findings are confirmed also in the case of mobile network, whose related analysis are presented in the Supplementary Information. Interestingly, however, in the case of mobile network we observe that over time, the correlation between wealth and speed showed a gradual decline. This phenomenon could be attributed to the higher demand in economically disadvantaged regions for more affordable connectivity options, such as mobile connection. Consequently, the evolution of mobile Internet may have diverged from that of fixed Internet due to distinct demands and consumer segmentation.

The present study comes with limitations. First, we used data from Internet Speedtest results provided by Ookla, which is only a proxy for Internet speed. As discussed, due to several factors, the outcome of tests might differ from the real Internet. Nevertheless, Ookla is widely used by academic and official institutions to measure Internet connectivity. Additionally, our methodology aims at attenuating some of the possible issues deriving from the heterogeneous use of this service, as detailed in Sec.~\ref{sec:mem} and in Ref.~\cite{gozzi2023adoption}. Second, we use only proxy data to measure wealth. Indeed, we consider the Relative Wealth Index published by Meta~\cite{chi2022microestimates} to characterize wealth at the desired spatial granularity, nonetheless such data come with inherent limitations, as is the case with all proxy measures. Lastly, Internet speed and wealth are linked by a feedback loop that we do not fully characterize due to data availability. As a result, our study mostly focuses on associations over time and space rather than causation or providing comprehensive explanations of the current landscape. 

Since $2020$, about $28$ USD billion have been invested in the telecom sector in Brazil~\cite{investmentBR}. Despite the significant amount of resources, the underlying efforts were not enough to provide a level playing field for all Internet users. This study, indeed, has shown how the poorest segments of population still experience a slower Internet connectivity compared to the most wealthy and how this gap may widen in case of exogenous shocks. Such disparity can have a significant impact on the socio-economic development of the country and requires a joint work of policy makers and the private sector to be solved. Specific policies at the local level should be promoted to improve connectivity in the poorest areas of towns, favoring the penetration of fiber to the Home (FTTH) technology, the affordability of high-speed Internet packages and devices, the development of specific digital skills through dedicated training and awareness programs. All these measures will support a more equal access to the Internet, ensuring that all individuals have access to a fast, affordable, and reliable Internet connection.


\section{Materials and methods}
\label{sec:mem}

\subsection{Measuring Internet speed}
We characterize Internet quality using as proxy Speedtest Intelligence\textsuperscript{\tiny\textregistered} data by Ookla~\cite{ookla}. Speedtest apps offer free analyses of Internet performance metrics. The tests are geolocalized and provide download/upload speed (expressed in Megabits per second). Here, following a common practice, we consider download speed as a metric to assess the quality of Internet. The dataset includes nearly $100M$ tests performed between 2017 and 2023, divided as follows: $47.5M$ in São Paulo, $24.1M$ in Rio de Janeiro, $8.1M$ in Belo Horizonte, $6.5M$ in Brasília, $6.1M$ in Fortaleza, and $4.9M$ in Manaus. 

We preprocess the data by excluding all tests displaying a download speed of $0$ Mbps, as these typically represent failed tests and do not provide informative insights into the actual network quality. Additionally, to limit the impact of outliers, we filter out tests with a download speed  $>2$ Gigabits per second, as this threshold is regarded the maximum value for broadband technology. After preprocessing, we compute Internet speed following a procedure similar to the one presented in Ref.~\cite{gozzi2023adoption}. 

To compute Internet speed in a geographical area $g$ over a timeframe $(t_1, t_2)$, we gather all tests conducted within that area during that period. Then, we calculate the median of the results obtained from tests conducted by individual users. In other words, for each user $u$, we compute the associated download speed as follows:
\begin{equation*}
    Mbps^{u, g}_{(t_1, t_2)} = med_i(Mbps^{u, g}_{i, (t_1, t_2)})
\end{equation*}

This step is taken to prevent bias caused by users who utilize the service more frequently than others. Finally, the download speed associated to area $g$ in timeframe $(t_1, t_2)$ is calculated as the median download speed across all users:

\begin{equation*}
    Mbps^{g}_{(t_1, t_2)} = med_u(Mbps^{u, g}_{(t_1, t_2)})
\end{equation*}

\subsection{Measuring wealth}
We assess the socio-economic status of different geographical regions using the Relative Wealth Index (RWI) from Meta's Data for Good Program~\cite{chi2022microestimates}. This index, made publicly available in 2021, offers micro-estimates of the relative standard of living within countries. It is built considering non-traditional data sources such as satellite imagery and privacy-preserving Facebook connectivity data, and it is validated by Meta through ground truth measurements obtained from the Demographic and Health Surveys. The RWI covers approximately $93$ low and middle-income countries globally, providing data at a high spatial resolution ($2.4km^2$ micro-regions). In this study, we aggregate the RWI at the desired geographical resolution by computing the average RWI of all micro-regions contained in the considered geography.

\subsection{Hexagonal grid}
We partition the area of each city considered into an hexagonal grid. This allows us to obtain a regular uniform spatial grid. We consider a resolution such that hexagonal units have an area of approximately $5.2km^2$. Figure~\ref{fig:fig5}A illustrates the resulting hexagonal grid for Manaus.

\subsection{Voronoi tessellation}

\begin{figure}[ht!]
\centering
\includegraphics[width=\textwidth]{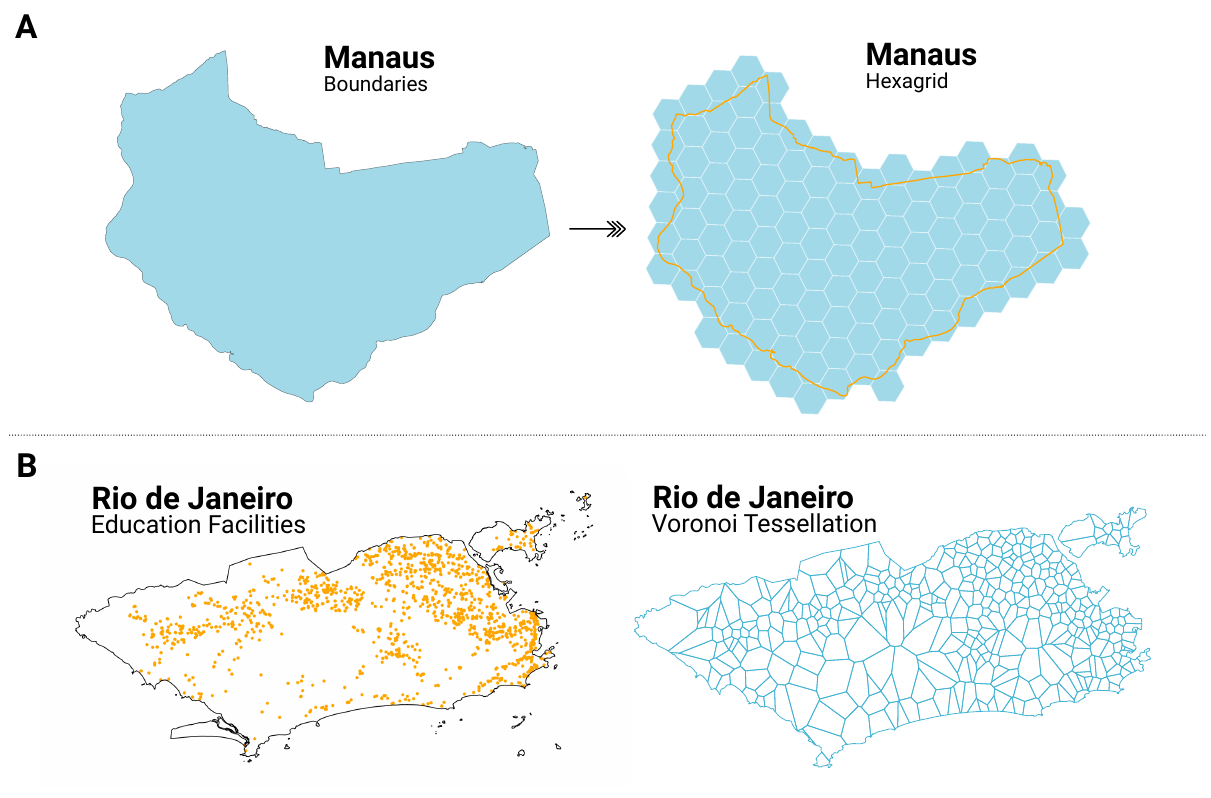}
\caption{\textbf{A)} Boundaries of Manaus and obtained hexagonal grid. \textbf{B)} Location of education facilities and obtained Voronoi tessellation for Rio de Janeiro.}
\label{fig:fig5}
\end{figure}

We collect the location data of educational facilities within the six cities under investigation from OpenStreetMap~\cite{hostm_edu}. We include all entities categorized with the tag ``amenity=school." Each facility is then condensed to its centroid, so that all facilities are represented by a unique set of coordinates. To prevent excessive fragmentation, we merge facilities located within a $1km$ radius of each other. Subsequently, we employ Voronoi tessellation on the resulting centroids. This process generates a Voronoi cell for each centroid, including all points on the plane that are closer to that seed point than to any other. This approach allows us to define the catchment areas of each educational facility.
Figure~\ref{fig:fig5}B illustrates the location of education facilities and the resulting Voronoi tessellation for Rio de Janeiro.

\section*{Acknowledgements}
This report was supported by the Digital Development Partnership, which aims to advance digital transformation in low and middle-income countries by building strong digital foundations and accelerators, facilitating digital use cases for the digital economy to thrive. All authors thank Ookla, The World Bank and the Development Data Partnership. All authors thank James Carroll, Katherine Macdonald, and 
Luciano Charlita De Freitas for their support and review.

\section{Supporting Information}

\subsection{Internet speed in the \textit{favelas} of Rio de Janeiro}
In the case of Rio de Janeiro, we extend our analysis to include tests conducted both inside and outside \textit{favelas}. \textit{Favelas} are informal, densely populated urban settlements in Brazil, typically characterized by substandard housing and a lack of basic services, arising from socio-economic disparities and rapid urbanization. For this analysis, we considered all tests performed inside and outside \textit{favelas} without prior aggregation into hexagonal units. The shapefile containing the boundaries of \textit{favelas} in Rio de Janeiro is sourced from Ref.~\cite{favelas_shp}. 

\begin{figure}[ht!]
\centering
\includegraphics[width=\textwidth]{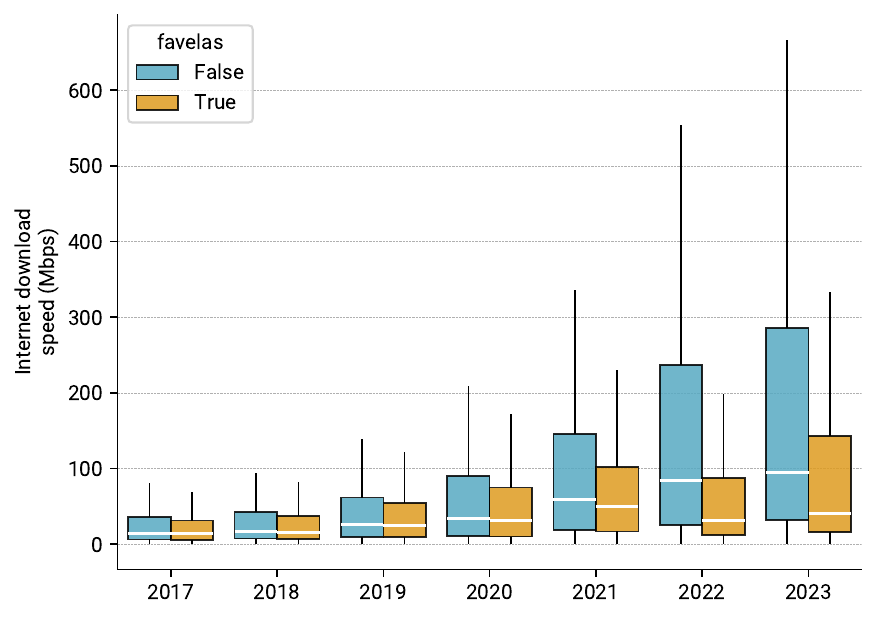}
\caption{Evolution of \textbf{fixed} Internet speed, expressed in download speed (Mbps), inside and outside the \textit{favelas} of Rio de Janeiro. Boxplots show Internet speed of each speed test performed witin and outside of \textit{favelas}.}
\label{fig:speed_favelas}
\end{figure}

Figure \ref{fig:speed_favelas} shows the evolution of internet speeds for tests performed inside and outside \textit{favelas}. Unsurprisingly, tests conducted within \textit{favelas} generally exhibit lower internet speeds. Additionally, this disparity has increased over the years. In $2017$, the median speeds for tests conducted inside and outside \textit{favelas} were $13.7$ Mbps and $14.3$ Mbps, respectively, reflecting a $4\%$ difference. By $2023$, these speeds had changed to $94.2$ Mbps and $40.1$ Mbps, respectively, resulting in a $57.4\%$ difference.

\subsection{Mobile Internet}
We repeat here the analysis presented in the main text considering mobile Internet speed instead of fixed.\\

\begin{figure}[ht!]
\centering
\includegraphics[width=\textwidth]{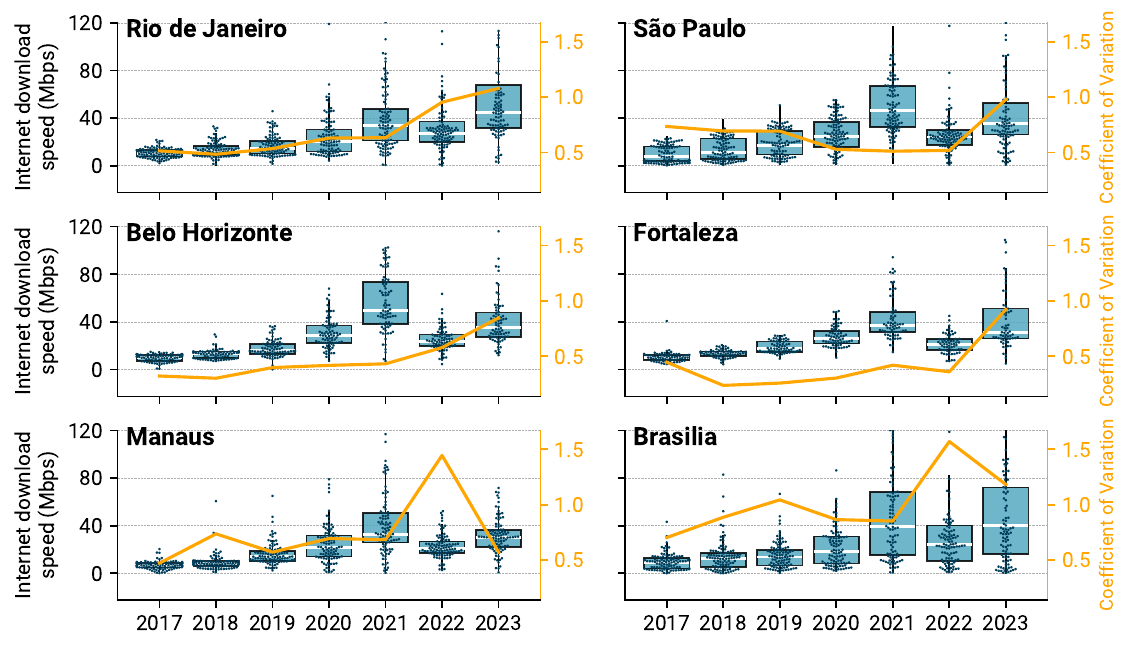}
\caption{Evolution of \textbf{mobile} Internet speed, expressed in download speed (Mbps), across the six cities considered. Boxplots show the distribution of Internet speed within each hexagonal unit in each city. The orange line indicates the evolution of the coefficient of variation of the Internet speed distribution over the years.}
\label{fig:speed_mobile}
\end{figure}

Figure~\ref{fig:speed_mobile} shows the evolution of mobile Internet speed, from $2017$ to $2023$ in the six cities. Across the board, our analysis reveals a significant improvement also in mobile Internet speed throughout all cities over the past six years. Specifically, Rio de Janeiro exhibits the highest median mobile download speed ($45mbps$) in $2023$, followed by Brasília ($40mpbs$), São Paulo and Belo Horizonte ($35mpbs$), Fortaleza ($31mpbs$), and finally Manaus ($30mpbs$). On the other hand, Manaus experienced the highest growth during the period, marking a $+348\%$ increase, followed closely by São Paulo ($+343\%$) Rio de Janeiro ($+337\%$) and Brasília ($+333\%$), and at a higher gap by Belo Horizonte ($+236\%$) and Fortaleza ($+211\%$). In the same plot we show the coefficient of variation of the distribution of mobile Internet speed within each city across the years. Our findings indicate an increasing trend in the coefficient of variation index across the six cities, suggesting a trend towards a more disperse distribution of mobile Internet speed as it improved. However, we acknowledge differences among the cities examined. This contrasts the finding obtained when considering fixed Internet speed, where we observed a trend towards more homogeneous distribution. Brasília exhibits the highest dispersion in mobile Internet speed distribution in $2023$ ($CV=1.20$), while Manaus features the most homogeneous distribution ($CV=0.57$).

\begin{figure}[ht!]
\centering
\includegraphics[width=\textwidth]{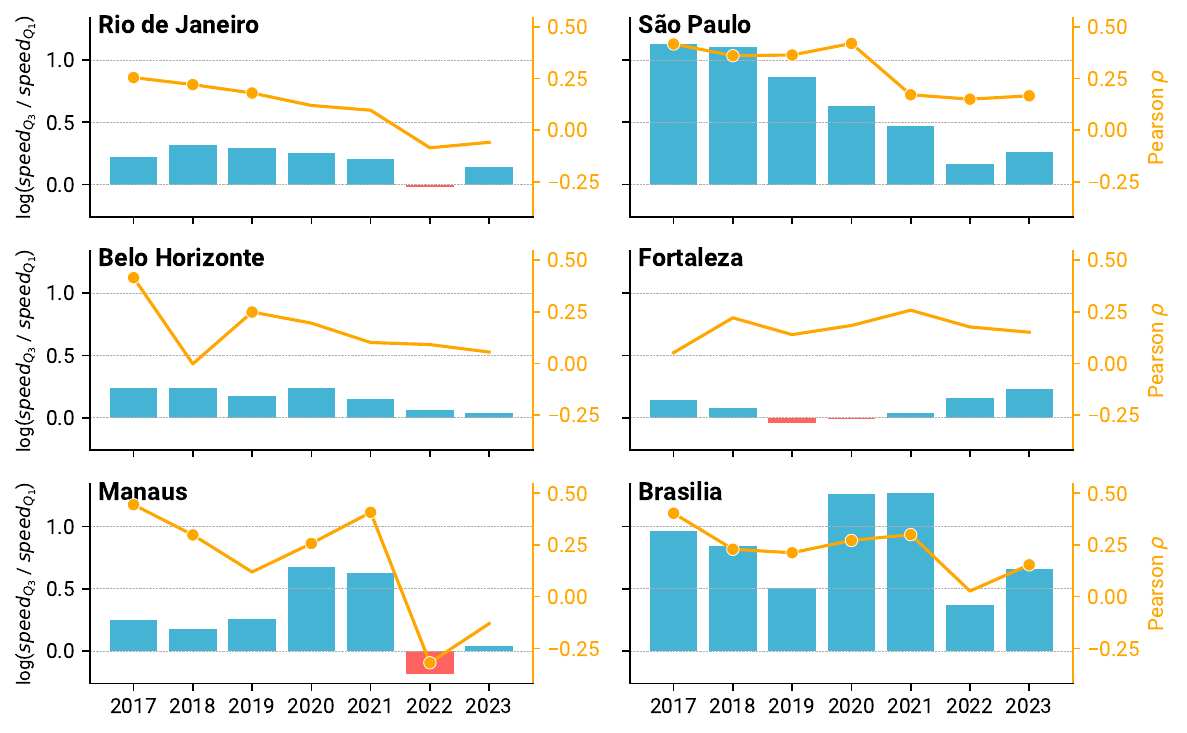}
\caption{Logarithm of the ratio between \textbf{mobile} Internet speeds measured in spatial units with wealth higher than the $75^{th}$ quantile and those with wealth lower than the $25^{th}$ quantile. The orange line represents the Pearson correlation coefficient between \textbf{mobile} Internet speed and RWI of different spatial units. Circles indicate where the coefficient is significant at the $5\%$ level.}
\label{fig:speed_ineq_mobile}
\end{figure}

Figure~\ref{fig:speed_ineq_mobile} shows the logarithm of the ratio between the average mobile Internet speed measured in cells with wealth higher than the $75^{th}$ quantile and those with wealth lower than the $25^{th}$ quantile. This metric is meant to compare and highlight the differences between the wealthiest and the poorest units. A value close to zero indicates similar Internet quality for both wealthy and less wealthy areas, while positive (negative) values denote better Internet quality for the more wealthy (less wealthy). 
Similarly to what found with fixed Internet speed, across various years and cities wealthy areas generally experience better mobile Internet quality compared to less wealthy areas. On the contrary, however, we note a dicreasing disparity in Internet quality between more and less wealthy areas over the years. Indeed, also the Pearson correlation coefficient between Internet speed and RWI tends to show a decreasing trend.

\begin{figure}[ht!]
\centering
\includegraphics[width=\textwidth]{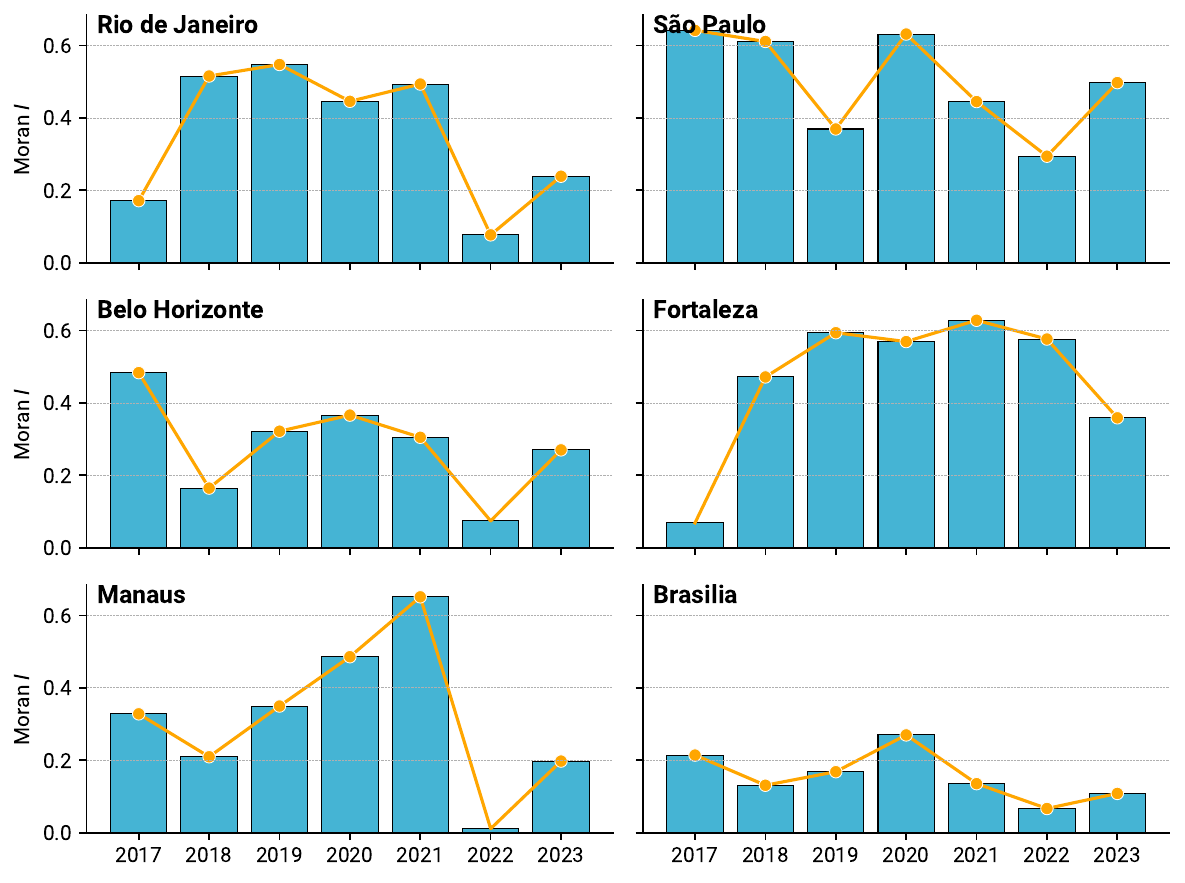}
\caption{\textbf{Spatial Clustering of Internet Mobile Speed.} Evolution of global Moran's $I$ in each city between $2017$ and $2023$. Circles indicate where the statistic is significant at the $5\%$ level.}
\label{fig:spatial_autocorr_mobile}
\end{figure}

In Figure~\ref{fig:spatial_autocorr_mobile} we show the evolution of the Moran's $I$ index of mobile Internet speed across different cities and years. We observe positive and significant indices in all cities, indicating spatial autocorrelation also of mobile Internet speed (i.e., areas with high mobile speed tend to be closer to areas also with high mobile Internet). Nonetheless, in the case of mobile Internet speed, the trend over the years is less clear. Overall, we notice a decrease in spatial autocorrelation, however this trend is not consistent and shared by all cities. 

\begin{figure}[ht!]
\centering
\includegraphics[width=\textwidth]{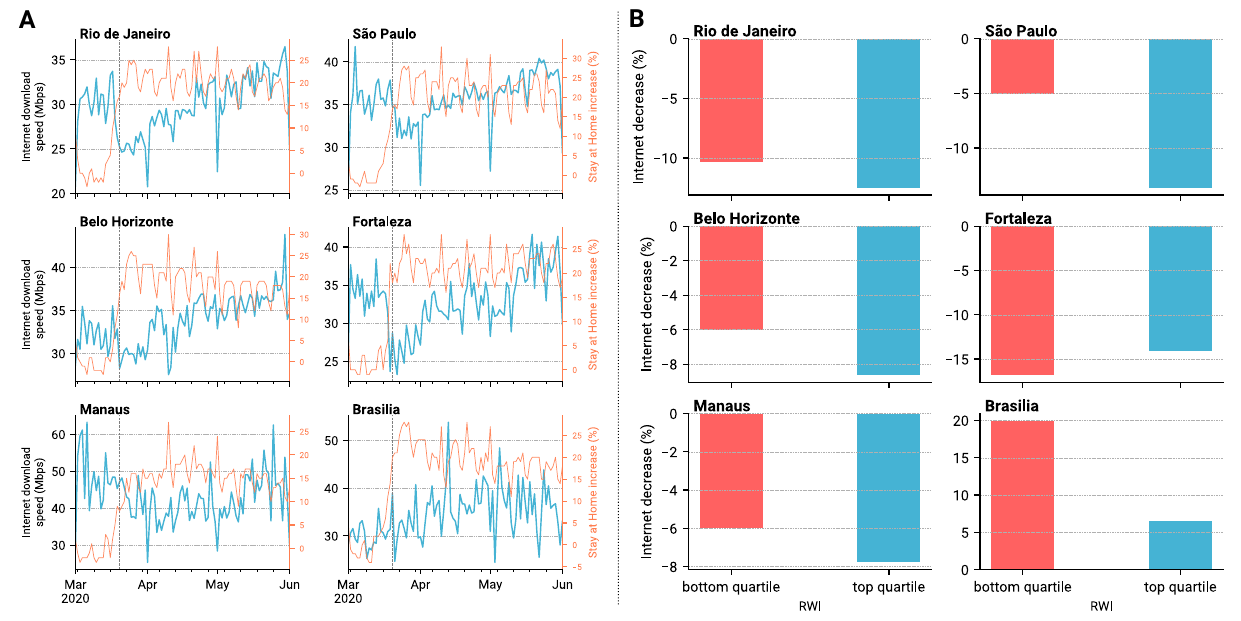}
\caption{\textbf{Mobile Network Resilience During the COVID-19 Pandemic}. \textbf{A)} Daily median \textbf{mobile} download speed in the six cities between March and June $2020$. The vertical dashed line indicates when Brazil declared the national emergency. The percentage change in individuals staying at home as measured via Google Community Mobility Reports is also shown. \textbf{B)} Drop in \textbf{mobile} Internet speed following the national emergency declaration in the top and bottom quartiles of the RWI in each city.}
\label{fig:covid_mobile}
\end{figure}

In Figure~\ref{fig:covid_mobile}A, we present the median daily mobile download speeds in the six cities between March and June $2020$. Additionally, we mark the date when Brazil declared a national emergency with a vertical dashed line and we show the increase in the percentage of individuals staying at home measured using data from the COVID-19 Community Mobility Reports published by Google~\cite{googleMob}. Across all cities, we observe a sharp decline in mobile network quality, as measured by download speed, following the declaration of the national emergency. Concurrently, the fraction of population staying at home increased. After the initial drop, we observe a gradual recovery, with mobile download speeds approaching pre-emergency levels by June $2020$. Among the cities considered, Fortaleza experienced the most significant drop in median mobile download speed computed in periods March $1^{st}$-March $20^{th}$ and  March $20^{th}$-April $1^{st}$, with a decline of $-20\%$, while Brasília is the only city showing a slight increase in speed $+3\%$. All other cities experienced declines ranging from Rio de Janeiro ($-17\%$),  São Paulo ($-9\%$), Manaus ($-8\%$), to Belo Horizonte ($-6\%$). Furthermore, in Figure~\ref{fig:covid_mobile}B, we illustrate these drops for the top and bottom quartiles of the RWI. We obtain a different picture from fixed Internet speed. Foraleza is the only cities where less wealthy areas experienced larger drop in mobile speed respect to more wealthy areas. In all other cities we observe an opposite trend.

\begin{figure}[ht!]
\centering
\includegraphics[width=\textwidth]{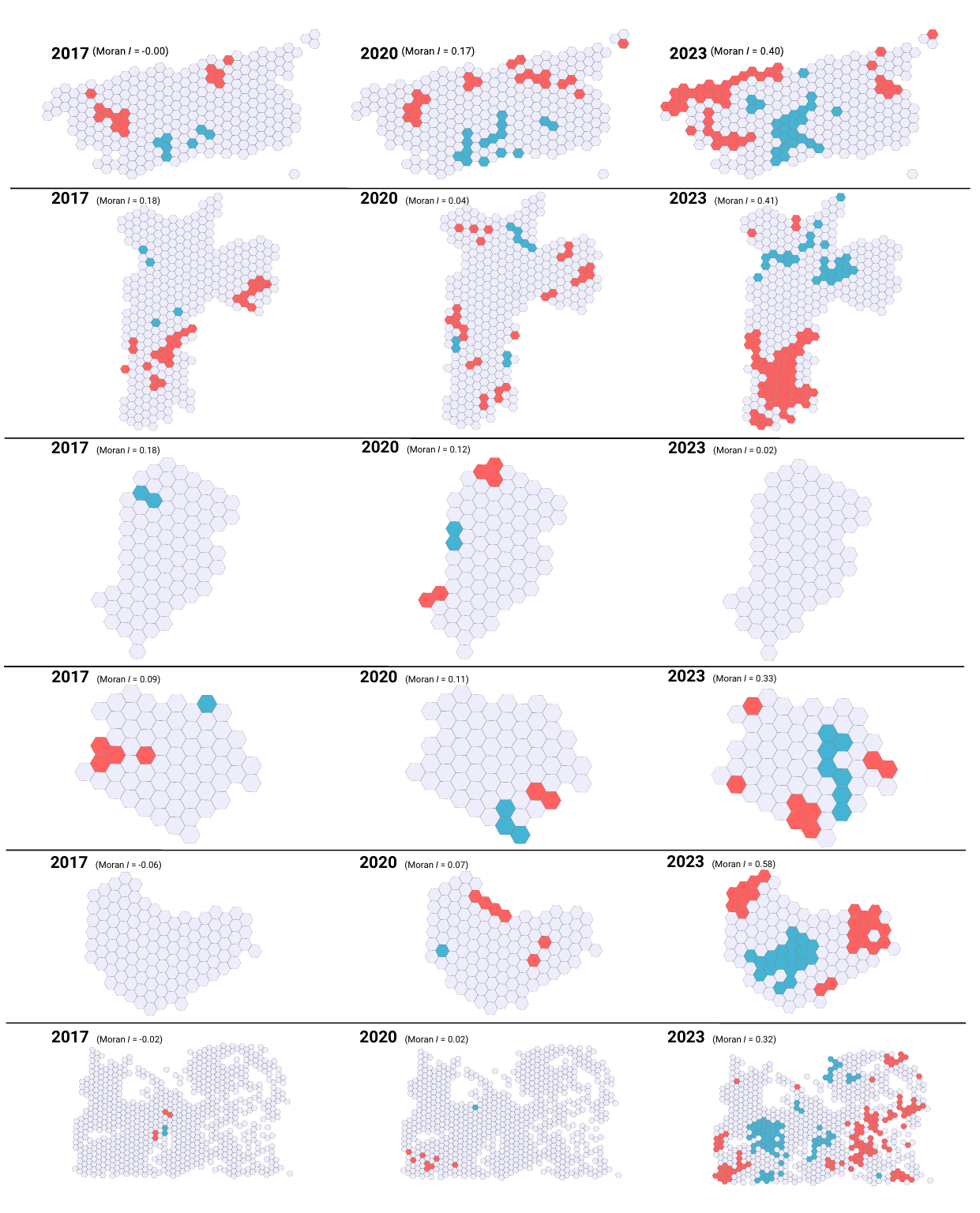}
\caption{\textbf{Spatial Clustering of Fixed Internet Speed.} Distribution of spatial units with significant local Moran's $I$ in all cities in $2017$, $2020$, and $2023$. Clusters of low (high) Internet speed are shown in red (blue).}
\label{fig:SIfig1}
\end{figure}

\clearpage

\end{document}